\makeatletter
\@ifundefined{@parse@version@dash}{%
\def\@parse@version#1{\@parse@version@0#1}
\def\@parse@version@#1/#2/#3#4#5\@nil{%
\@parse@version@dash#1-#2-#3#4\@nil}
\def\@parse@version@dash#1-#2-#3#4#5\@nil{%
  \if\relax#2\relax\else#1\fi#2#3#4 }
}{}
\makeatother

\documentclass[%
 preprint,
 titlepage,
 superscriptaddress,
 amsmath,
 amssymb,
 aps,
 pra,
 showkeys,
 showpacs,
]{revtex4-2}
\usepackage{graphicx}
\usepackage{dcolumn}
\usepackage{bm}
\usepackage{hyperref}

\begin{document}
\preprint{APS/123-QED}
\title{Decoherence Effect of Qubits in 1D Transverse Ising Model}

\author{Bobin Li}
\thanks{Corresponding author:gslibobin@lut.edu.cn}%
\affiliation{Department of Physics, Lanzhou University of Technology, Lanzhou 730050, China}
\date{\today}
\begin{abstract}
With the development of lab technology, the low-order correlation function can no longer describe the main effect of decoherence in quantum many-body system, so it is imperative to study the higher-order correlation function of the system. In this paper, we study the changes of the correlation functions in the decoherence effect, analytically. And explore when it is possible to approach the qubit decoherence process only by low-order correlation function, and when third-order or higher correlation functions are needed in 1D transverse Ising model. It indicates that, under strong coupling and long coherence time, the effect of high-order correlation functions can not be ignored, and the approximation of classical Markov process is limited. But, in the case of weak coupling and short coherence time, low-order correlation function can describe well.
\end{abstract}

\maketitle

\section{Introduction}

The decoherence of a qubit is caused by its bath, which is essentially a quantum many-body system. Previously, people studied the decoherence of qubits, and the decoherence effect of the bath is generally approached by the classical Markov process, which is described by the second-order correlation function. Under this approximation, the effect of higher-order correlation function are ignored. In recent years, with the development of quantum control and dynamic decoupling technology, the coherent time of qubits has been greatly extended, so the effect of quantum many body increasingly stand out, then the markov approximation cannot satisfy the needs of the quantum many body system, and, namely, the second-order correlation function cannot cover the main effect of baths, and the papers$^{[1]}$ can obtain strict solutions and comprehensive properties of higher order correlation functions for many-body systems.

For example, it can be solved strictly to 1D Ising model and 2D Ising model with zero magnetic field by analytic method. It is available for thosed systems with not very strong many-body interaction$^{[2,3,4]}$  by using more advanced numerical methods, such as CCE so on. And the numerical solution$^{[5]}$ of higher order correlation function is obtained. For systems with strong many-body interaction but small numbers of particles(less than 30 particles), they can be solved by the method of numerically strict diagonalization, citing Haiqing Lin $^{[5]}$. For systems with strong many-body interaction and large numbers of particles, it will be available by the methods of extrapolation that is to extrapolate from the weak interaction strength and less particle numbers to those cases that is strong interaction and large numbers, and it could obtain some properties of higher order correlation function qualitatively.

At the same time, with the update of experimental technology, it becomes feasible for the experimental measurement of high-order correlation function , which has promoted the theoretical improvement for the research of correlation function, and the study of high-order correlation function is imperative. Therefore, this work mainly focuses on the role of higher-order correlation function in the process of decoherence of qubit. By theoretical analysis for many-body quantum effect in complex baths, it is obtained strictly analytically for the form of higher order correlation function. The contribution of the high-order correlation function is proposed by comparing with the exact solution in 1D transverse Ising model. It is found that in the process of strong coupling and long time interaction between system and bath, the contribution of high-order correlation function will increase for to the decoherence of qubit.

\section{Theory method}
\subsection{Correlation function}

The reduced density matrix of the central spin Qubit $\rho^{S}(t)$ is
\begin{align*}
\rho^{S}(t) &=Tr_{B}\rho(t)=Tr_{B}(\mathcal{T}exp(\int_{0}^{t}d\tau\mathcal{L}(\tau))\rho(0))\\
 & =\sum_{n}\frac{1}{n!}Tr_{B}(\int_{0}^{t}\mathcal{T}[\mathcal{L}(t_{1})\mathcal{L}(t_{2})\cdots\mathcal{L}(t_{n})\rho(0)]dt_{1}dt_{2}\cdots dt_{n})\\
& =\sum_{n}\frac{2^{n}}{n!}\sum_{\{\alpha_{n}\}}\sum_{\{\eta_{n}\}}\int_{0}^{t}dt_{1}\cdots dt_{n}\cdot C_{\alpha_{n}\cdots\alpha_{1}}^{\eta_{n}\cdots\eta_{1}}\times(\mathcal{T}[\mathcal{S}_{\alpha_{1}}^{\overline{\eta}_{1}}(t_{1})\cdots\mathcal{S}_{\alpha_{n}}^{\overline{\eta}_{n}}(t_{n})]\rho^{S}(0))
\end{align*}
where $\mathcal{L}(t)=2\sum_{\alpha}(\mathcal{S}_{\alpha}^{+}(t)\mathcal{B}_{\alpha}^{-}(t)+\mathcal{S}_{\alpha}^{-}(t)\mathcal{B}_{\alpha}^{+}(t))$
So $C_{\alpha_{n}\cdots\alpha_{1}}^{\eta_{n}\cdots\eta_{1}}=Tr_{B}(T\mathcal{B}_{\alpha_{1}}^{\eta_{1}}(t_{1})\cdots\mathcal{B}_{\alpha_{n}}^{\eta_{n}}(t_{n})\rho^{B})$ Is the correlation function, while the irreducible correlation function is $\tilde{C}_{\alpha_{n}\cdots\alpha_{1}}^{\eta_{n}\cdots\eta_{1}}$[8],
The system dynamics can be described as
\begin{align}
\rho^{S}(t) =\mathcal{T}\exp(\sum_{N=1}^{\infty}\frac{2^{N}}{N!}\int_{0}^{t}dt_{1}\cdots dt_{n}\cdot\tilde{C}_{\alpha_{n}\cdots\alpha_{1}}^{\eta_{n}\cdots\eta_{1}}\times(\mathcal{S}_{\alpha_{n}}^{\overline{\eta}_{n}}(t_{n})\cdots\mathcal{S}_{\alpha_{1}}^{\overline{\eta}_{1}}(t_{1}))\rho^{S}(0) \end{align}

\subsection{1D Transverse Ising model}
The bath is regarded as a 1D transverse Ising model. In this model, Surrounded by a circle of bath spins, the qubit is located on the central axis of the circle. In this bath, the bath spins are coupled in only one direction, and there is an external magnetic field in the whole environment. The exact solution of 1D transverse Ising model is shown in Appendix A. Since it is applied to be periodic boundary conditions, qubit is located in the center of the circle of 1D transverse Ising model chain, and the system Hamiltonian in the Schrodinger representation is:
\begin{align}
H=H_{0}+V=\omega_{0}\sigma^{z}-\sum_{j=1}^{N}\sigma_{j}^{x}\sigma_{j+1}^{x}-\lambda\sum_{j=1}^{N}\sigma_{j}^{z}+\sum_{j=1}^{N}\sigma^{z}\otimes(-g\sigma_{j}^{z})
\end{align}
where $H_{0}=\omega_{0}\sigma^{z}-\sum_{j=1}^{N}\sigma_{j}^{x}\sigma_{j+1}^{x}-\lambda\sum_{j=1}^{N}\sigma_{j}^{z}$and
$V=\sum_{j=1}^{N}\sigma^{z}\otimes(-g\sigma_{j}^{z})$.
Use the Jordan--Wigner transform
\[
\sigma_{j}^{z}=1-2a_{j}^{\dagger}a_{j},\sigma_{j}^{x}+i\sigma_{j}^{y}=2(\prod_{i<j}\sigma_{i}^{z})a_{j}
\]
The system Hamiltonian was changed to a Fermi system
\[
H_{\lambda}=-\sum_{j=1}^{N}[(a_{j}^{\dagger}-a_{j})(a_{j+1}^{\dagger}+a_{j+1})-2\lambda a_{j}^{\dagger}a_{j}]-\lambda N+\omega_{0}\sigma^{z}-g\sum_{j=1}^{N}\sigma^{z}\otimes(1-2a_{j}^{\dagger}a_{j})
\]
By the Fourier transform $a_{j}=\sum_{k}c_{k}exp(-ikj)\sqrt{N}$,The spin system is mapped to the spin-free Fermi system
\[
H_{\lambda}=-\sum_{k}[(2cosk-2\lambda)c_{k}^{\dagger}c_{k}+isink(c_{-k}^{\dagger}c_{k}^{\dagger}+c_{-k}c_{k})]-N\lambda+\omega_{0}\sigma^{z}-g\sum_{k}\sigma^{z}\otimes(1-2c_{k}^{\dagger}c_{k})
\]
where$c_{k}^{\dagger}c_{k}$ is the creation annihilation operator of fermions with wave vector K. The system Hamiltonian can be diagonally transformed by the Bogoliubov transformation
\[
\begin{bmatrix}b_{-k}\\
b_{k}\\
b_{-k}^{\dagger}\\
b_{k}^{\dagger}
\end{bmatrix}=\begin{bmatrix}u_{k} & 0 & 0 & iv_{k}\\
0 & u_{k} & -iv_{k} & 0\\
0 & -iv_{k} & u_{k} & 0\\
iv_{k} & 0 & 0 & u_{k}
\end{bmatrix}\begin{bmatrix}c_{-k}\\
c_{k}\\
c_{-k}^{\dagger}\\
c_{k}^{\dagger}
\end{bmatrix}
\]
Here, $u_{k}=cos\theta_{k},v_{k}=sin\theta_{k}$with $tan(2\theta_{k})=sink/(cosk-\lambda)$. After the transformation, the diagonalized Fermionic Hamiltonian is
\begin{align}
H_{\lambda}=\sum_{k}\varepsilon_{k}(b_{k}^{\dagger}b_{k}-1/2)+\omega_{0}\sigma^{z}-g\sum_{k}\sigma^{z}\otimes(cos2\theta_{k}-2b_{k}^{\dagger}b_{k}-isin2\theta_{k}(b_{k}^{\dagger}b_{-k}^{\dagger}-b_{-k}b_{k})))
=H+V
\end{align}
The interaction representation, the interaction Hamiltonian$V(t)$
\begin{align}
V(t)=(-g\sigma^{z})\otimes\sum_{k}(cos2\theta_{k}-2b_{k}^{\dagger}b_{k} -isin2\theta_{k}\cdot(b_{k}^{\dagger}b_{-k}^{\dagger}\cdot exp(-i2\varepsilon_{k}t)-b_{-k}b_{k}\cdot exp(i2\varepsilon_{k}t)))=S_{1}B_{1}
\end{align}
At the same times
\begin{align}
S_{1} & =-g\sigma^{z}\\
B_{1} & =\sum_{k}(cos2\theta_{k}-2b_{k}^{\dagger}b_{k}-isin2\theta_{k}\cdot(b_{k}^{\dagger}b_{-k}^{\dagger}\cdot exp(-i2\varepsilon_{k}t)-b_{-k}b_{k}\cdot exp(i2\varepsilon_{k}t)))
\end{align}
where $u_{k}=cos\theta_{k},v_{k}=sin\theta_{k}$with $tan(2\theta_{k})=sink/(cosk-\lambda)$
where $\varepsilon_{k}=2\sqrt{1-2\lambda cosk+\lambda^{2}}$

\section{Results and Discussion}

For the density matrix of the system, there is an initial state relation
\[
\rho(0)=\rho^{S}(0)\otimes\rho^{B}(0)
\]
Where bath is $\rho^{B}(0)=\frac{1}{Z}e^{-\beta H_{0}}$,The density matrix of the qubit
\[
\rho_{10}^{S}(t)=\rho_{01}^{S*}(t)=\rho_{10}^{S}(0)e^{\Gamma(t)}
\]
where $\Gamma(t)$ Is the decoherence function. Since 1D transverse Field Ising model is a pure dephasing model, so
\begin{align*}
\rho^{S}(t) & =exp[\sum_{N=0}^{+\infty}\frac{2^{N}}{N!}\int_{0}^{t}dt_{N}\cdots dt_{1}\tilde{C}_{1\cdots1}^{+\cdots+}\mathcal{S}_{1}^{-}\cdots\mathcal{S}_{1}^{-}]\rho^{S}(0)\\
 & =\prod_{N=0}^{\infty}\sum_{n=0}^{+\infty}(\frac{1}{n!}[\frac{2^{N}}{N!}\int_{0}^{t}dt_{N}\cdots dt_{1}\tilde{C}_{1\cdots1}^{+\cdots+}]^{n}[ig]^{nN})\left[\begin{array}{cc}
0 & c\\
(-1)^{nN}\cdot c & 0
\end{array}\right]
\end{align*}
So
\begin{align*}
\rho_{10}^{S}(t)=exp[\sum_{N=0}^{+\infty}(\frac{2^{N}}{N!}\int_{0}^{t}dt_{N}\cdots dt_{1}\tilde{C}_{1\cdots1}^{+\cdots+}[ig]^{N})]\cdot c
\end{align*}
So the series representation of the decoherence function is obtained as follow
\begin{align}
\Gamma(t)=\sum_{N=0}^{+\infty}(\frac{2^{N}}{N!}\int_{0}^{t}dt_{N}\cdots dt_{1}\tilde{C}_{1\cdots1}^{+\cdots+}[ig]^{N})
\end{align}

\paragraph{The first-order correlation function}
\begin{align*}
C_{1}^{+} =\sum_{k}(cos2\theta_{k}-\frac{2}{e^{\beta\varepsilon_{k}}+1})
\end{align*}
The first-order irreducible correlation function
\begin{align}
C_{1}^{+} =\tilde{C}_{1}^{+}=\sum_{k}(cos2\theta_{k}-\frac{2}{e^{\beta\varepsilon_{k}}+1})\stackrel{\beta\rightarrow\infty}{=}\sum_{k}cos2\theta_{k}
\end{align}

\paragraph{The second-order correlation function}
\begin{align*}
C_{11}^{++} & =\sum_{k,k'}(cos2\theta_{k}(cos2\theta_{k'}-\frac{2}{e^{\beta\varepsilon_{k'}}+1})-\frac{2}{e^{\beta\varepsilon_{k}}+1}(cos2\theta_{k'}-\frac{2}{e^{\beta\varepsilon_{k'}}+1}))\\
 & +\sum_{k}(sin^{2}2\theta_{k}(cos(2\varepsilon_{k}(t_{1}-t_{2}))\frac{1}{(e^{\beta\varepsilon_{k}}+1)^{2}}+cos(2\varepsilon_{k}(t_{1}-t_{2}))(\frac{1}{e^{\beta\varepsilon_{k}}+1}+1)^{2})\\
 & \stackrel{\beta\rightarrow\infty}{=}\sum_{k,k'}cos2\theta_{k}cos2\theta_{k'}+\sum_{k}cos(2\varepsilon_{k}(t_{1}-t_{2}))
\end{align*}
The second-order irreducible correlation function
\begin{align}
\tilde{C}_{11}^{++} =\sum_{k}cos(2\varepsilon_{k}(t_{1}-t_{2}))(\frac{1}{e^{\beta\varepsilon_{k}}+1}+1)^{2}
 \stackrel{\beta\rightarrow\infty}{=}\sum_{k}cos(2\varepsilon_{k}(t_{1}-t_{2}))
\end{align}

\paragraph{The third-order correlation function}
\begin{align}
\tilde{C}_{111}^{+++}
 &\stackrel{\beta\rightarrow\infty}{=}-\sum_{k_{1}}sin^{2}2\theta_{k_{1}}[[1-\theta(t_{3}-t_{1})\theta(t_{1}-t_{2})-\theta(t_{1}-t_{3})\theta(t_{3}-t_{2})]cos(2\varepsilon_{k_{1}}(t_{1}-t_{3}))\nonumber
 \\ & +[1-\theta(t_{2}-t_{1})\theta(t_{1}-t_{3})-\theta(t_{1}-t_{2})\theta(t_{2}-t_{3})]cos(2\varepsilon_{k_{1}}(t_{1}-t_{2}))\nonumber \\ & +[1-\theta(t_{3}-t_{2})\theta(t_{2}-t_{1})-\theta(t_{2}-t_{3})\theta(t_{3}-t_{1})]cos(2\varepsilon_{k_{1}}(t_{2}-t_{3}))]
\end{align}
where $tan(2\theta_{k})=sink/(cosk-\lambda)$ and $\varepsilon_{k}=2\sqrt{1-2\lambda cosk+\lambda^{2}}$,
Applied $sin(2\theta_{k})=sink\text{/\ensuremath{\sqrt{1-2\lambda cosk+\lambda^{2}}}}$,
$cos(2\theta_{k})=(cosk-\lambda)/\sqrt{1-2\lambda cosk+\lambda^{2}}$.
See Appendix B for the derivation of the third-order correlation function in detail.

\centerline{\includegraphics[width=13.0cm]{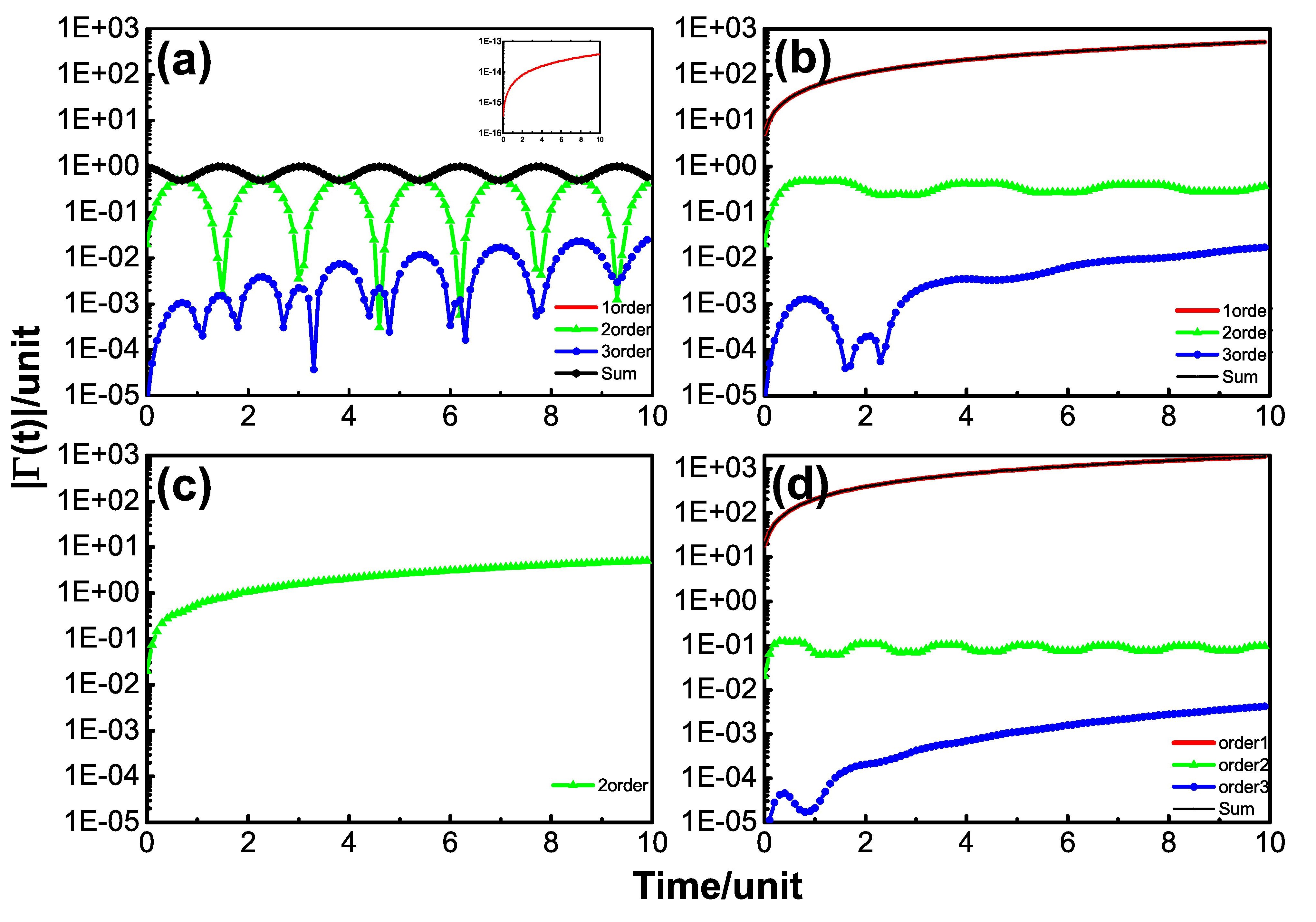}}
\vskip 2mm
\centerline{\parbox[c]{14cm}}{\footnotesize Fig.~1.~ $N=10000,g=0.01$, Under weak coupling,
the first three order correlation function changes with the external field intensity. (a)$\lambda=0.0$,(b)$\lambda=0.5$,(c)phase transition point$\lambda=1.0$,(d)$\lambda=2.0$}
\vskip 0.55\baselineskip 

Firstly, qubit is approximated to the weak coupling of $G =0.01$in the thermodynamic limit. 
With the increase of external magnetic field $\lambda$, the bath system transfers from ferromagnetic phase$0.0\leqslant \lambda<1.0$ to paramagnetic phase $\lambda>1.0$. In $\lambda=1.0$,the system underwent a quantum phase transition.
As shown in Figure 1(c), at this phase point, the total correlation function of the system diverges, because that the correlation length of the system tends to infinity. Specifically, the odd-order of the obtained correlation function diverges, and only the second-order correlation function remains limited. In the weak field condition, the first two order correlation functions shows the feasibility of Markov approximation that are always larger than the third-order, with time evolving.

The proportion of the third order correlation function is increasing with time, but it always contributes less to the total correlation function.It can be foreseen that, with the evolution time continues, it will become more and more important for the proportion of the correlation function above the second order. Therefore, when the time evolution of the decoherence process is relatively long, The Markov approximation faces the dilemma of too large accumulation error. And this change is independent of the phase in which the system is located, as shown in FIG. 1(b) and 1(d).

\centerline{\includegraphics[width=13.0cm]{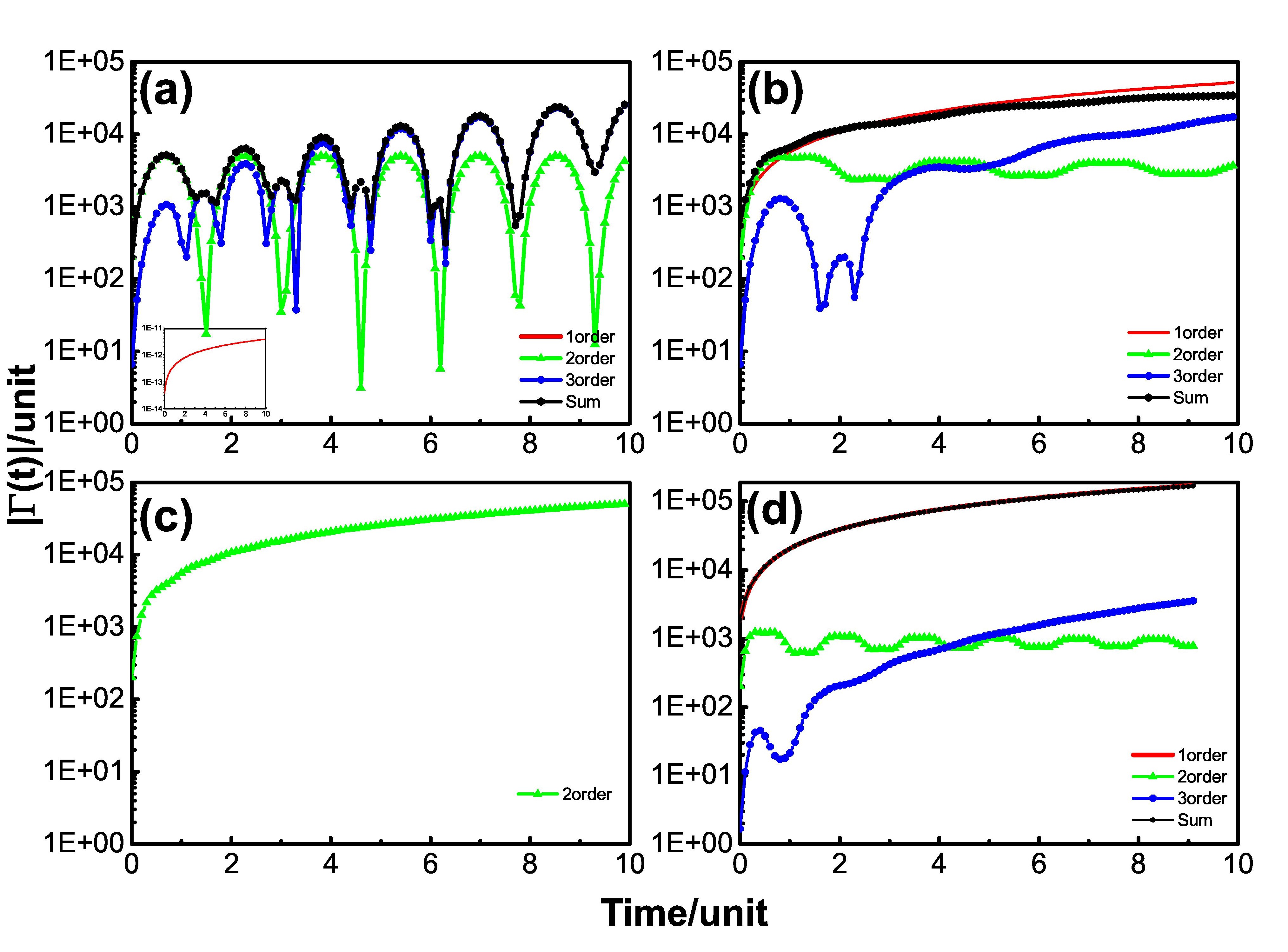}}
\centerline{\parbox[c]{14cm}}{\footnotesize Fig.~2.~$N=10000,g=1.0$,Under strong coupling,
the first three order correlation function changes with the external field intensity. (a)$\lambda=0.0$,(b)$\lambda=0.5$,(c)phase transition point$\lambda=1.0$,(d)$\lambda=2.0$}
\vskip 0.55\baselineskip
Under the same parameters, the coupling strength of the system is increased to $g=1.0$ ,becoming a strongly coupled system. FIG. 2 is consistent with FIG. 1 for the phase transformation process of the system. However, the difference is that the third-order correlation function, that is, the non-Markov term, exceeds the second-order correlation function in a relatively short time. This shows that in the case of strong coupling, non-Markov terms will become important.

\centerline{\includegraphics[width=13.0cm]{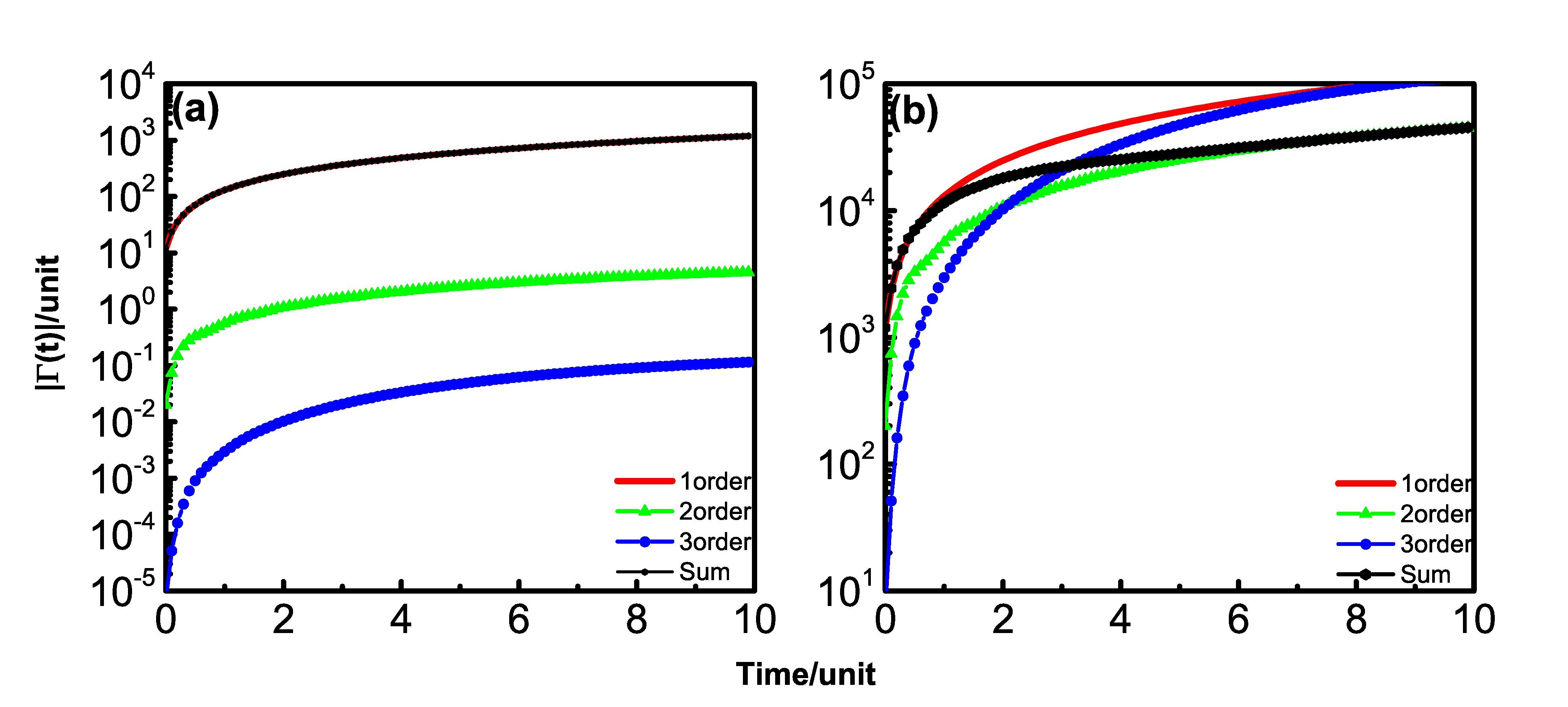}}
\centerline{\parbox[c]{14cm}}{\footnotesize Fig.~3.~ $N=10000,\lambda=0.97$,Near phase transition point$\lambda=1.0$,The first three order changes.(a)$g=0.01$,(b)$g=1.0$}
\vskip 0.55\baselineskip

\centerline{\includegraphics[width=13.0cm]{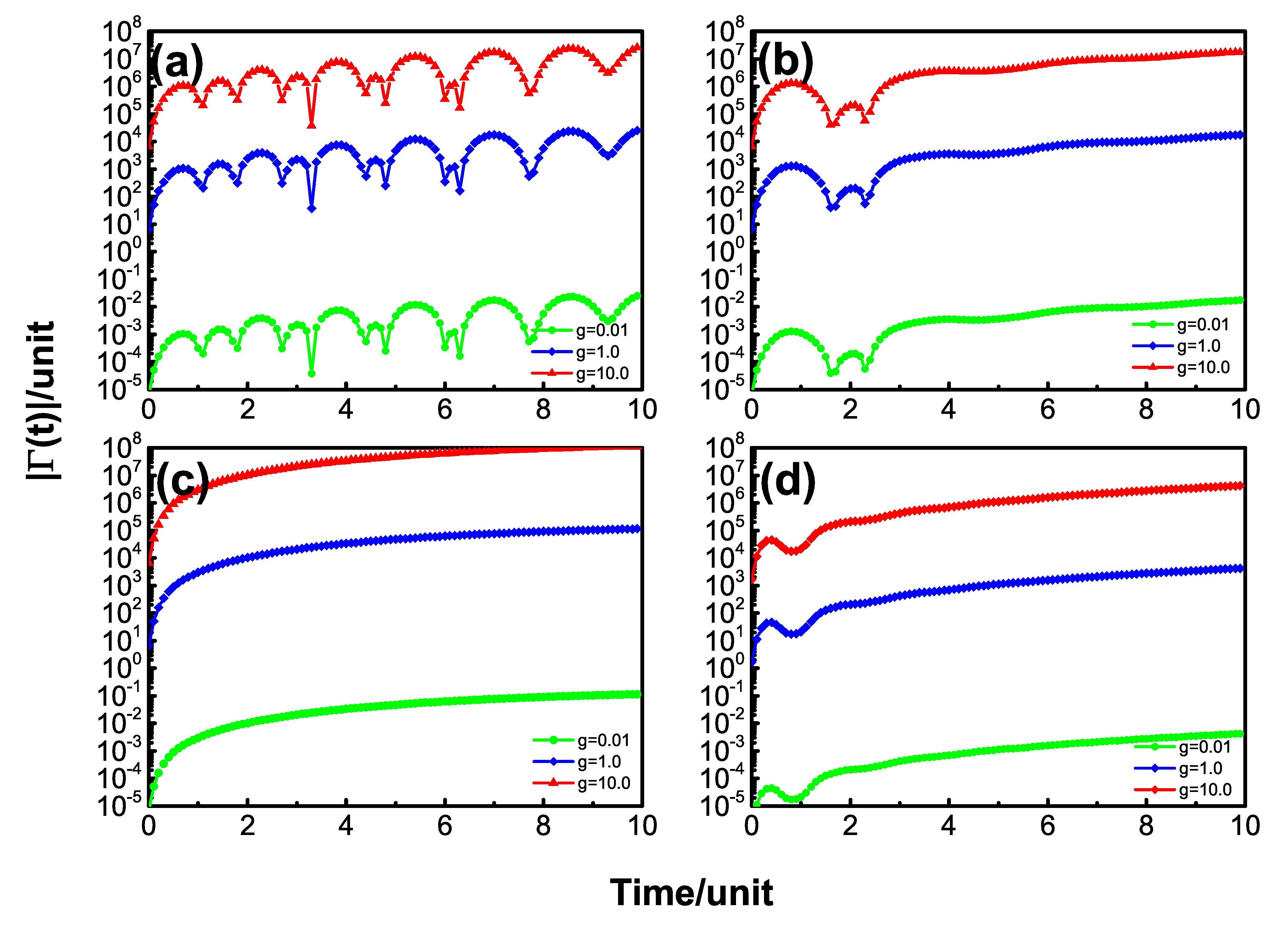}}
\centerline{\parbox[c]{14cm}}{\footnotesize Fig.~4.~ $N=10000$,The three order changes.(a)$\lambda=0.0$,(b)$\lambda=0.5$,(c)$\lambda=0.97$,(d)$\lambda=2.0$}
\vskip 0.55\baselineskip

Since the correlation function diverges at the phase transition point $\lambda=1.0$, it is difficult to explore the effect of phase transition on the system. So in FIG. 3, the state of the system immediately adjacent to the phase transition point $\lambda=0.97$is applied. First, as before, non-Markov terms become important as the coupling strength increases. Moreover, when the phase transition point is approached, the quantum interference effect disappears, showing a smooth relationship, indicating that the phase transition is a statistical effect rather than a quantum effect.

In FIG. 4, it can be clearly seen that with the increase of coupling strength $G $of the system, non-Markov term modes become larger and larger, and this growth almost linearly increases with the increase of coupling strength. When the system is in different phase states, the non-Markov terms have different trends with the change of coherent time. In ferromagnetic and paramagnetic states, far from the phase transition point, the non-Markov term has quantum interference effect. There is a monotone increase as we approach the phase transition.

\section{Conclusion}
We study the effect of bath on the decoherence of qubit, which mainly refers to the contribution of the high-order correlation function of 1D transverse Ising model.The larger the coupling strength of the quantum many-body system, the more obvious the effect of the higher order correlation function.The longer coherent time of the qubit in the quantum many-body bath, the more obvious the effect of the higher order correlation function.The closer the quantum many-body system is to the phase transition point, the more obvious the effect of the higher order correlation function is.At the phase transition point, the higher-order correlation function should also mutate.

\newpage
\appendix
\section{The exact solution of the transverse field Ising model}
\begin{align}
H =H_{S}+H_{b}+g\sigma_{Z}B =|\uparrow\rangle\langle\uparrow|\otimes H_{+}+|\downarrow\rangle\langle\downarrow|\otimes H_{-}
\end{align}
Where$H_{+}=(\omega_{0}+H_{b}+gB)$ and $H_{-}=(-\omega_{0}+H_{b}-gB)$
At the same times,
\begin{align}
\rho(0)=\rho^{S}(0)\otimes\rho^{B}(0)and\rho^{B}(0)=\frac{1}{Z}e^{-\beta H_{0}}
\end{align}
Secondly
\begin{align}
\rho^{S}(t)=tr_{b}(\rho(t))=tr_{b}(U(t,0)\rho(0)U^{+}(t,0))
\end{align}
Finally
\begin{align}
U(t,0)=e^{-itH}=|\uparrow\rangle\langle\uparrow|e^{-itH_{+}}+|\downarrow\rangle\langle\downarrow|e^{-itH_{-}}
\end{align}
So
\begin{align}
\rho_{\downarrow\uparrow}^{S}(t) =\langle\uparrow|tr_{b}(\rho(t))|\downarrow\rangle=c\cdot tr_{b}[e^{-itH_{+}}\frac{1}{Z}e^{-\beta H_{b}}e^{itH_{-}}]
\end{align}
when temperature is zero $\beta\rightarrow\infty$,we have
\begin{align}
\rho_{\downarrow\uparrow}^{S}(t) & =\langle\uparrow|tr_{b}(\rho(t))|\downarrow\rangle\\ \nonumber
 & =c\cdot e^{-it2\omega_{0}}\langle g|e^{-it(H_{b}+gB)}e^{it(H_{b}-gB)}|g\rangle
\end{align}
The bases of Hilbert space
\begin{align}
\{\tilde{|0_{-k},0_{k}\rangle},\tilde{|1_{-k},1_{k}\rangle},\tilde{|0_{-k},1_{k}\rangle},\tilde{|1_{-k},0_{k}\rangle}\}
\end{align}
Under those bases
\begin{align*}
H_{b} & =\left[\begin{array}{cccc}
-\frac{1}{2}\sum_{k'}\varepsilon_{k'} & 0 & 0 & 0\\
0 & -\frac{1}{2}\sum_{k'}(-1)^{\delta_{k',\pm k}}\varepsilon_{k'} & 0 & 0\\
0 & 0 & -\frac{1}{2}\sum_{k'}(1-\delta_{k',\pm k})\varepsilon_{k'} & 0\\
0 & 0 & 0 & -\frac{1}{2}\sum_{k'}(1-\delta_{k',\pm k})\varepsilon_{k'}
\end{array}\right]\\
 & =H_{b}^{k}-\frac{1}{2}\sum_{k'\neq\pm k}\varepsilon_{k'}
\end{align*}
So
\begin{align*}
B & =\left[\begin{array}{cccc}
\sum_{k'}cos2\theta_{k'} & isin2\theta_{k}\cdot e^{i2\varepsilon_{k}t} & 0 & 0\\
-isin2\theta_{k}\cdot e^{-i2\varepsilon_{k}t} & \sum_{k'}cos2\theta_{k'}-4 & 0 & 0\\
0 & 0 & \sum_{k'}cos2\theta_{k'}-2 & 0\\
0 & 0 & 0 & \sum_{k'}cos2\theta_{k'}-2
\end{array}\right]\\
 & =B^{k}+\sum_{k'}cos2\theta_{k'}
\end{align*}
So
\begin{align*}
\rho_{\downarrow\uparrow}^{S}(t) & =\langle\uparrow|tr_{b}(\rho(t))|\downarrow\rangle\\
 & =c\cdot e^{-it2\omega_{0}}\langle g|e^{-it(H_{b}+gB)}e^{it(H_{b}-gB)}|g\rangle\\
 & =c\cdot e^{-it2(\omega_{0}+g\sum_{k'}cos2\theta_{k'})}\prod_{k>0}\langle g|U_{k}^{\dagger}(t)U_{k}(t)|g\rangle
\end{align*}
So
\begin{align*}
U_{k}^{\dagger}(t)U_{k}(t) & =e^{-it(H_{b}^{k}+gB^{k})}e^{it(H_{b}^{k}-gB^{k})}\\
 & =\left[\begin{array}{cc}
e^{-itM_{1}}\cdot e^{itM_{2}} & 0\\
0 & e^{i4tg}I
\end{array}\right]
\end{align*}
Where
\begin{align*}
e^{-itM_{1}}\cdot e^{itM_{2}} =e^{4itg}\cdot e^{-it\vec{a}\cdot\vec{\sigma}}\cdot e^{it\vec{b}\cdot\vec{\sigma}}
\end{align*}
where $\vec{a}=(-sin2\theta_{k}sin(2\varepsilon_{k}t)g,-sin2\theta_{k}cos(2\varepsilon_{k}t)g,(2g-\varepsilon_{k})),\\ \vec{b}=(sin2\theta_{k}sin(2\varepsilon_{k}t)g,sin2\theta_{k}cos(2\varepsilon_{k}t)g,(-2g-\varepsilon_{k}))$and
$\vec{\sigma}=(\sigma_{1},\sigma_{2},\sigma_{3})$ is vector of Puli matrix.

\begin{align*}
e^{-itM_{1}}\cdot e^{itM_{2}} & =e^{4itg}\cdot e^{-it\vec{a}\cdot\vec{\sigma}}\cdot e^{it\vec{b}\cdot\vec{\sigma}}\\
 & =e^{4itg}\cdot\left[\begin{array}{cc}
A & B\\
C & D
\end{array}\right]
\end{align*}
Where
\begin{align*}
A & =cos(ta)cos(tb)+(\varepsilon_{k}^{2}-sin^{2}2\theta_{k})\frac{sin(ta)sin(tb)}{ab}+i\varepsilon_{k}[\frac{sin(ta)cos(tb)}{a}-\frac{cos(ta)sin(tb)}{b}]\\
B & =sin2\theta_{k}\cdot e^{i2\varepsilon_{k}t}[\frac{cos(ta)sin(tb)}{b}+\frac{sin(ta)cos(tb)}{a}+i2(\varepsilon_{k}+2)\frac{sin(ta)sin(tb)}{ab}]\\
C & =-sin2\theta_{k}\cdot e^{-i2\varepsilon_{k}t}[\frac{sin(ta)cos(tb)}{a}+\frac{cos(ta)sin(tb)}{b}-i2(\varepsilon_{k}-2)\frac{sin(ta)sin(tb)}{ab}]\\
D & =cos(ta)cos(tb)+(\varepsilon_{k}^{2}-16-sin^{2}2\theta_{k})\frac{sin(ta)sin(tb)}{ab}+i[(\varepsilon_{k}+4)\frac{sin(tb)cos(ta)}{a}\\
& -(\varepsilon_{k}-4)\frac{cos(tb)sin(ta)}{b}]
\end{align*}
Where $a=\sqrt{(gsin2\theta_{k})^{2}+(2g-\varepsilon_{k})^{2}},b=\sqrt{(gsin2\theta_{k})^{2}+(2g+\varepsilon_{k})^{2}}$
As a same times
\begin{align}
\rho_{\downarrow\uparrow}^{S}(t)=e^{\Gamma(t)}\rho_{\downarrow\uparrow}^{S}(0)
\end{align}
So
\begin{align}
e^{\Gamma(t)}=\prod_{k>0}[cos(ta)cos(tb)+(\varepsilon_{k}^{2}-sin^{2}2\theta_{k})\frac{sin(ta)sin(tb)}{ab}+i\varepsilon_{k}[\frac{sin(ta)cos(tb)}{a}-\frac{cos(ta)sin(tb)}{b}]]
\end{align}
So
\begin{align}
\Gamma(t)=\sum_{k>0}In(cos(ta)cos(tb)+(\varepsilon_{k}^{2}-sin^{2}2\theta_{k})\frac{sin(ta)sin(tb)}{ab}+i\varepsilon_{k}[\frac{sin(ta)cos(tb)}{a}-\frac{cos(ta)sin(tb)}{b}]\text{)}
\end{align}
Where $a=\sqrt{(gsin2\theta_{k})^{2}+(2g-\varepsilon_{k})^{2}},b=\sqrt{(gsin2\theta_{k})^{2}+(2g+\varepsilon_{k})^{2}}$and
$\varepsilon_{k}=2\sqrt{1-2\lambda cosk+\lambda^{2}}$and $sin2\theta_{k}=sink\text{/\ensuremath{\sqrt{1-2\lambda cosk+\lambda^{2}}}}$and
$k=\frac{(2l-1)\pi}{N},l=1,2,\cdots\frac{N}{2}$

\section{The third order correlation function}
\begin{align*}
Part & =tr(B_{1}(t_{3})B_{1}(t_{2})B_{1}(t_{1})\rho^{B}(0))\\
 & =\sum_{k_{3}k_{2}k_{1}}(cos2\theta_{k_{3}}-2\frac{1}{e^{\beta\varepsilon_{k_{3}}}+1})(cos2\theta_{k_{2}}-2\frac{1}{e^{\beta\varepsilon_{k_{2}}}+1})(cos2\theta_{k_{1}}-2\frac{1}{e^{\beta\varepsilon_{k_{1}}}+1})\\
 & +\sum_{k_{3}k_{1}}(cos2\theta_{k_{3}}-2\frac{1}{e^{\beta\varepsilon_{k_{3}}}+1})(sin^{2}2\theta_{k_{1}}\cdot[\frac{1}{e^{\beta\varepsilon_{k_{1}}}+1}\frac{1}{e^{\beta\varepsilon_{-k_{1}}}+1}\cdot \\ &
  exp(i2\varepsilon_{k_{1}}(t_{1}-t_{2}))+(\frac{1}{e^{\beta\varepsilon_{k_{1}}}+1}+1)(\frac{1}{e^{\beta\varepsilon_{-k_{1}}}+1}+1)\cdot exp(-i2\varepsilon_{k_{1}}(t_{1}-t_{2}))])\\
  & +\sum_{k_{2}k_{1}}(sin^{2}2\theta_{k_{1}}cos2\theta_{k_{2}}\cdot\frac{1}{e^{\beta\varepsilon_{k_{1}}}+1}\frac{1}{e^{\beta\varepsilon_{-k_{1}}}+1}\cdot[exp(i2(\varepsilon_{k_{1}}(t_{1}-t_{3}))
 +exp(-i2\varepsilon_{k_{1}}(t_{1}-t_{3}))])  \\
 &+\sum_{k_{2}k_{1}}(-2sin^{2}2\theta_{k_{1}}\cdot((\frac{1}{e^{\beta\varepsilon_{k_{2}}}+1}-1)\cdot\frac{1}{e^{\beta\varepsilon_{-k_{1}}}+1}\frac{1}{e^{\beta\varepsilon_{k_{1}}}+1})\cdot exp(i2(\varepsilon_{k_{1}}(t_{1}-t_{3})))\\
 & +\sum_{k_{2}k_{1}}(sin^{2}2\theta_{k_{2}}\cdot[\frac{1}{e^{\beta\varepsilon_{k_{2}}}+1}\frac{1}{e^{\beta\varepsilon_{-k_{2}}}+1}\cdot exp(i2\varepsilon_{k_{2}}(t_{2}-t_{3}))+(\frac{1}{e^{\beta\varepsilon_{k_{2}}}+1}+1)(\frac{1}{e^{\beta\varepsilon_{-k_{2}}}+1}+1)\cdot \\ & exp(-i2\varepsilon_{k_{2}}(t_{2}-t_{3}))]\cdot(cos2\theta_{k_{1}}-2\frac{1}{e^{\beta\varepsilon_{k_{1}}}+1})\\
 & +\sum_{k_{2}k_{1}}(-2sin^{2}2\theta_{k_{1}}\cdot(1+\frac{1}{e^{\beta\varepsilon_{k_{2}}}+1})(1+\frac{1}{e^{\beta\varepsilon_{k_{1}}}+1})(1+\frac{1}{e^{\beta\varepsilon_{-k_{1}}}+1})\cdot exp(-i2(\varepsilon_{k_{1}}(t_{1}-t_{3})))\\
 & \stackrel{\beta\rightarrow\infty}{=}\sum_{k_{3}k_{2}k_{1}}[cos2\theta_{k_{3}}cos2\theta_{k_{2}}cos2\theta_{k_{1}}+cos2\theta_{k_{3}}sin^{2}2\theta_{k_{1}}\cdot \\ & exp(-i2\varepsilon_{k_{1}}(t_{1}-t_{2}))+cos2\theta_{k_{1}}sin^{2}2\theta_{k_{2}}\cdot exp(-i2\varepsilon_{k_{2}}(t_{2}-t_{3}))-2sin^{2}2\theta_{k_{1}}\cdot exp(-i2(\varepsilon_{k_{1}}(t_{1}-t_{3}))]
\end{align*}

\begin{align}
Part\stackrel{\beta\rightarrow\infty}{=} & \sum_{k_{3}k_{2}k_{1}}[cos2\theta_{k_{3}}cos2\theta_{k_{2}}cos2\theta_{k_{1}}+cos2\theta_{k_{3}}sin^{2}2\theta_{k_{1}}\cdot exp(-i2\varepsilon_{k_{1}}(t_{1}-t_{2}))\nonumber   \\
& +cos2\theta_{k_{1}}sin^{2}2\theta_{k_{2}}\cdot exp(-i2\varepsilon_{k_{2}}(t_{2}-t_{3}))-2sin^{2}2\theta_{k_{1}}\cdot exp(-i2(\varepsilon_{k_{1}}(t_{1}-t_{3}))]
\end{align}

\newpage
\begin{align*}
 & C_{111}^{+++}\\
 & =\frac{1}{2^{2}}[[1-\theta(t_{3}-t_{1})\theta(t_{1}-t_{2})-\theta(t_{1}-t_{3})\theta(t_{3}-t_{2})][Part+Part(t_{1}\leftrightarrow t_{3})]\\
 & +[1-\theta(t_{2}-t_{1})\theta(t_{1}-t_{3})-\theta(t_{1}-t_{2})\theta(t_{2}-t_{3})][Part(t_{1}\rightarrow t_{2},t_{2}\rightarrow t_{3},t_{3}\rightarrow t_{1})+Part(t_{2}\leftrightarrow t_{3})]\\
 & +[1-\theta(t_{3}-t_{2})\theta(t_{2}-t_{1})-\theta(t_{2}-t_{3})\theta(t_{3}-t_{1})][Part(t_{2}\leftrightarrow t_{1})+Part(t_{1}\rightarrow t_{3},t_{2}\rightarrow t_{1},t_{3}\rightarrow t_{2})]]\\
 & \overset{\beta\rightarrow\infty}{=}\frac{1}{2^{2}}[[1-\theta(t_{3}-t_{1})\theta(t_{1}-t_{2})-\theta(t_{1}-t_{3})\theta(t_{3}-t_{2})]\times[\sum_{k_{3}k_{2}k_{1}}[2\cdot\\
 &cos2\theta_{k_{3}}cos2\theta_{k_{2}}cos2\theta_{k_{1}}+cos2\theta_{k_{3}}sin^{2}2\theta_{k_{1}}\cdot(exp(i2\varepsilon_{k_{1}}(t_{2}-t_{1}))+exp(i2\varepsilon_{k_{1}}(t_{2}-t_{3})))\\
 &+cos2\theta_{k_{1}}sin^{2}2\theta_{k_{2}}\cdot(exp(i2\varepsilon_{k_{2}}(t_{3}-t_{2}))+exp(i2\varepsilon_{k_{2}}(t_{1}-t_{2})))-2sin^{2}2\theta_{k_{1}}\cdot2cos(2\varepsilon_{k_{1}}(t_{1}-t_{3}))]\\
 & +[1-\theta(t_{2}-t_{1})\theta(t_{1}-t_{3})-\theta(t_{1}-t_{2})\theta(t_{2}-t_{3})]\\
 & \times[\sum_{k_{3}k_{2}k_{1}}[2\cdot cos2\theta_{k_{3}}cos2\theta_{k_{2}}cos2\theta_{k_{1}}+cos2\theta_{k_{3}}sin^{2}2\theta_{k_{1}}\cdot(exp(i2\varepsilon_{k_{1}}(t_{3}-t_{2}))+exp(i2\varepsilon_{k_{1}}(t_{3}-t_{1})))\\
 &+cos2\theta_{k_{1}}sin^{2}2\theta_{k_{2}}\cdot(exp(i2\varepsilon_{k_{2}}(t_{1}-t_{3}))+exp(i2\varepsilon_{k_{2}}(t_{2}-t_{3})))-2sin^{2}2\theta_{k_{1}}\cdot2cos(2\varepsilon_{k_{1}}(t_{1}-t_{2}))]\\
 & +[1-\theta(t_{3}-t_{2})\theta(t_{2}-t_{1})-\theta(t_{2}-t_{3})\theta(t_{3}-t_{1})]\\
 & \times[\sum_{k_{3}k_{2}k_{1}}[2\cdot cos2\theta_{k_{3}}cos2\theta_{k_{2}}cos2\theta_{k_{1}}+cos2\theta_{k_{3}}sin^{2}2\theta_{k_{1}}\cdot(exp(i2\varepsilon_{k_{1}}(t_{1}-t_{2}))+exp(i2\varepsilon_{k_{1}}(t_{1}-t_{3})))\\
 &+cos2\theta_{k_{1}}sin^{2}2\theta_{k_{2}}\cdot(exp(i2\varepsilon_{k_{2}}(t_{3}-t_{1}))+exp(i2\varepsilon_{k_{2}}(t_{2}-t_{1})))-2sin^{2}2\theta_{k_{1}}\cdot2cos(2\varepsilon_{k_{1}}(t_{2}-t_{3}))]]
\end{align*}

where $tan(2\theta_{k})=sink/(cosk-\lambda)$ and $\varepsilon_{k}=2\sqrt{1-2\lambda cosk+\lambda^{2}}$,
Applied $sin(2\theta_{k})=sink\text{/\ensuremath{\sqrt{1-2\lambda cosk+\lambda^{2}}}}$,
$cos(2\theta_{k})=(cosk-\lambda)/\sqrt{1-2\lambda cosk+\lambda^{2}}$.

The irreducible correlation function as follow
\begin{align*}
\tilde{C}_{111}^{+++} & =\frac{1}{2^{2}}[[1-\theta(t_{3}-t_{1})\theta(t_{1}-t_{2})-\theta(t_{1}-t_{3})\theta(t_{3}-t_{2})][\sum_{k_{1}}-2sin^{2}2\theta_{k_{1}}\cdot2cos(2\varepsilon_{k_{1}}(t_{1}-t_{3}))]\\
 & +[1-\theta(t_{2}-t_{1})\theta(t_{1}-t_{3})-\theta(t_{1}-t_{2})\theta(t_{2}-t_{3})][\sum_{k_{1}}-2sin^{2}2\theta_{k_{1}}\cdot2cos(2\varepsilon_{k_{1}}(t_{1}-t_{2}))]\\
 & +[1-\theta(t_{3}-t_{2})\theta(t_{2}-t_{1})-\theta(t_{2}-t_{3})\theta(t_{3}-t_{1})][\sum_{k_{1}}-2sin^{2}2\theta_{k_{1}}\cdot2cos(2\varepsilon_{k_{1}}(t_{2}-t_{3}))]]\\
 & =-\sum_{k_{1}}sin^{2}2\theta_{k_{1}}[[1-\theta(t_{3}-t_{1})\theta(t_{1}-t_{2})-\theta(t_{1}-t_{3})\theta(t_{3}-t_{2})]cos(2\varepsilon_{k_{1}}(t_{1}-t_{3}))\\
 & +[1-\theta(t_{2}-t_{1})\theta(t_{1}-t_{3})-\theta(t_{1}-t_{2})\theta(t_{2}-t_{3})]cos(2\varepsilon_{k_{1}}(t_{1}-t_{2}))\\
 & +[1-\theta(t_{3}-t_{2})\theta(t_{2}-t_{1})-\theta(t_{2}-t_{3})\theta(t_{3}-t_{1})]cos(2\varepsilon_{k_{1}}(t_{2}-t_{3}))]
\end{align*}
So

\begin{align}
\tilde{C}_{111}^{+++} &
=-\sum_{k_{1}}sin^{2}2\theta_{k_{1}}[[1-\theta(t_{3}-t_{1})\theta(t_{1}-t_{2})-\theta(t_{1}-t_{3})\theta(t_{3}-t_{2})]cos(2\varepsilon_{k_{1}}(t_{1}-t_{3})) \nonumber \\
 & +[1-\theta(t_{2}-t_{1})\theta(t_{1}-t_{3})-\theta(t_{1}-t_{2})\theta(t_{2}-t_{3})]cos(2\varepsilon_{k_{1}}(t_{1}-t_{2})) \nonumber \\
 & +[1-\theta(t_{3}-t_{2})\theta(t_{2}-t_{1})-\theta(t_{2}-t_{3})\theta(t_{3}-t_{1})]cos(2\varepsilon_{k_{1}}(t_{2}-t_{3}))]
\end{align}
\newpage
\bibliography{apssamp}

\begin{thebibliography}{999}
\bibitem{1} G. Gasbarri and L. Ferialdi.Stochastic unravelings of non-Markovian completely positive and trace-preserving maps. PhysRevA.98.042111
\bibitem{2} Wen Yang and Ren-Bao Liu.Quantum many-body theory of qubit decoherence in a finite-size spin bath. PhysRevB.78.085315
\bibitem{3} Wang Yao, Ren-Bao Liu, and L. J. Sham. Theory of electron spin decoherence by interacting nuclear spins in a quantum dot. PhysRevB.74.195301(2006)
\bibitem{4} W. M. Witzel and S. Das Sarma. Quantum theory for electron spin decoherence induced by nuclear spin dynamics in semiconductor quantum computer architectures: Spectral diffusion of localized electron spins in the nuclear solid-state environment. PhysRevB.74.035322(2006)
\bibitem{5} Yan-Chao Li and Hai-Qing Lin. Thermal quantum and classical correlations and entanglement in the XY spin model with three-spin interaction. PhysRevA.83.052323
\end{thebibliography}

\end{document}